# Optical imaging of transient acoustic fields using a phase contrast method★

G.T. Clement and S.V. Letcher
Department of Physics, University of Rhode Island, Kingston
Email: gclement@bwh.harvard.edu

A coherent phase-contrast optical system has been designed and tested for tomographic imaging of pressure fields from experimental transient acoustic signals. The system is similar to the pulsed, central-order schlieren method, but uses a Fourier filtering technique that images the actual acoustic pressure field, where the former technique reconstructs only the absolute value of the field. Simulations of the system are performed using a single-cycle sine-wave acoustic pulse. Experimental images resulting from a broad-band Gaussian pulse input to an underwater piezoceramic transducer array are presented. Relative pressure field s are reconstructed in space over a series of times after the source excitation. Theory and limitations of the phase contrast system are discussed.

# I. Introduction

Optical imaging techniques have been used extensively to reconstruct pressure fields radiating from acoustic transmitters. Acoustooptic diffraction theory[1,2] describes the light-sound interaction that permits this imaging. These virtually noninvasive techniques offer both high resolution and sensitivity. In particular, the pulsed schlieren optical technique[3] has the ability to record rapidly an acoustic field over finite distances at a specific time. This technique has been used to image both harmonic[3] and transient[4] fields. Experimental methods have been developed to allow imaging of signals with frequencies as low as 100 kHz.[5] Traditional schlieren systems, however, return the absolute value of a pressure field, creating ambiguity between regions of positive and negative pressure. As a result, the method is more commonly used for qualitative studies requiring the position of a wavefront at a specific time.

A phase-contrast optical method[6] is presented here, which returns data linearly proportional to the actual pressure field. In place of schlieren central-order or " knife-edge" filtering,[7] which serves as a cut off filter in the spatial frequency domain, an optical Hilbert transformer is used to shift and attenuate the zero-order of the diffracted light . The result is an amplitude-modulated projection of the sound field. The projection is tomographically reconstructed to give a three-dimensional representation of the sound field at a particular time. The theory motivating this method is similar to that of a phase-contrast microscope.[8] In Sec. 3, this theory is presented and compared with schlieren central-order filtering. This section also describes the tomographic reconstruction algorithm used to reconstruct the field in 3-D space at a given time. The acoustic field considered is generated by a circular planar source, thus an axisymmetric field propagating normal to the source is assumed. A Hankel-transform-based algorithm[9] exploit s the cylindrical symmetry of the field for efficient reconstruction.

The experimental apparatus and procedure are described in Sec 4. A detailed discussion of the phase-contrast filter is included where the specific criteria for the selected filter dimensions are explained. An overview of the optical setup and apparatus used is also presented. In Sec. 5 we present the results of a Gaussian-shaped pulse input into a cylindrically symmetric underwater transducer and the result s are compared with synthetic data obtained using an impulse-response method.

## II. Theory

The imaging theory in this paper considers a collimated coherent light source, assumed to be a plane wave of wavenumber $k = \omega / c_0$ in vacuum, traveling in the Cartesian $z$ direction. The planar light field,

$$E(z,t) = E_0 e^{i(\omega t - kz)}, \tag{1}$$

interacts with a small localized variation in optical refractive index $n$ $(x, y, z, t)$ due to sound field propagation. The interacting light pulse is assumed very short, so that the index of refraction may be considered at a constant time, $t_0$. After the light–sound interaction, the initially-planar electromagnetic field will experience a spatial phase variation given by

$$\Delta\phi(x,y) = 2k \int_0^R (n_0 - n(x,y,z,t_0))dz, \tag{2}$$

where $n_0$ is the unperturbed refractive index of water and $k_0$ is the optic wavenumber of the light source. The refractive index variation is directly proportional to pressure for small pressure change,[10] allowing Eq. (2) to be described by

$$\Delta\phi(x,y) = 2kC \int_0^R p(x,y,z)dz. \tag{3}$$

Immediately following the interaction, the resulting field becomes

$$E(x,y,z) = E_0 e^{i(\omega t - kz)} e^{i\Delta\phi(x,y)}.$$

(4)

It is assumed that *n* varies only slightly over a wavelength of light, λ, and that the phase shift Δφ(x,y) << 1 . Under the latter assumption, Δφ is expanded in a Taylor series keeping terms to the first order,

$$E = E_0 e^{i(\omega t - kz)}(1 + i\Delta\phi).$$

(5)

The equivalent result may also be obtained from the Raman-Nath equations[2] in the weak interaction limit.

The first term in Eq. (5) is simply the initial planar field. A converging lens or mirror in the optical far field returns the spatial transform of the field at its focal plane.[11] Thus, the undisturbed field appears in the center of this plane, while higher order terms are due to the phase shift. If a filter is placed in the center of the focal plane that shifts the phase of the zero order by π/2 and attenuates its field amplitude by a factor $\eta$, the reconstructed field in the image plane becomes

$$E = i E_0 e^{i(\omega t - kz)}(\eta + \Delta\phi),$$

(6)

where ideally $\eta \sim \varphi$ . The light intensity is recorded in the image plane; keeping only first order terms

$$\frac{I}{I_0} \simeq (\eta + \Delta\phi)^2,$$

(7)

showing the square-root of the intensity proportional to the phase shift .

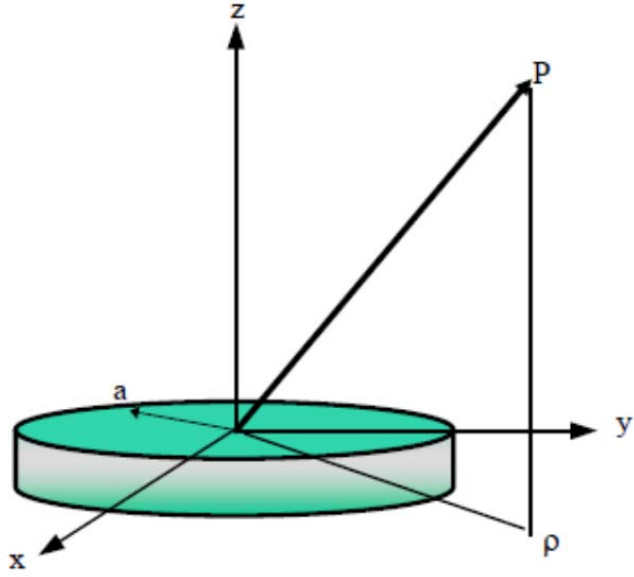

Figure 1. Axisymmetric radiator normal to the z-axis

The phase shift is related by Eq. (3) to the integral of the acoustic pressure $p$ along $z$. In practice, the full field $p(x, y, z, t_0)$ is determined using the tomographic reconstruction algorithm discussed below. The net effect of the filter is to transform a phase modulated (PM) image to an amplitude (AM) modulated one, allowing it to be easily recorded. In order to reconstruct the field, the following function is defined:

$$f(x, k_y, t_0) = \frac{1}{\sqrt{2\pi}} \int_{-\infty}^{\infty} \frac{\Delta\varphi(x, y, t_0)}{kC} e^{ik_y y} dy, \tag{8}$$

where $f$ is recognized as the spatial Fourier transform with respect to $y$ of Eq. (3). Substitution for $\Delta\varphi$ from Eq. (3) into Eq. (8) yields,

$$f(x, k_y, t_0) = \frac{1}{\sqrt{2\pi}} \int_{-\infty}^{\infty} \int_{-\infty}^{\infty} p(x, y, z, t_0) e^{ik_y y} dy dz. \tag{9}$$

The experimental sound source is shown in Fig. 1 to be an axisymmetric planar radiator,

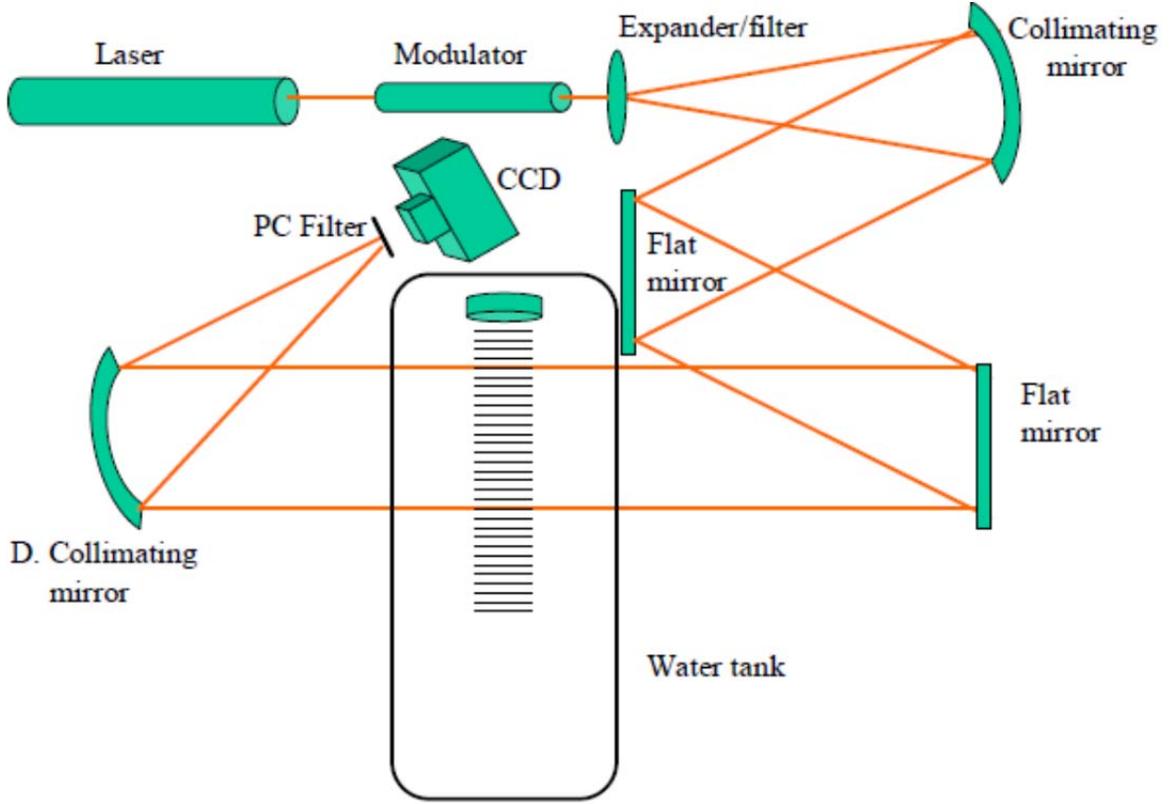

Figure 2. Phase-contrast imaging apparatus.

positioned normal to the x-axis. Thus, the index of refraction becomes $n(x, y, z, t_0)$ and the pressure field $p(x, y, z, t_0)$. Utilizing this symmetry, a change of variables to cylindrical coordinates is made,

$$f(x, k_y, t_0) = \frac{1}{\sqrt{2\pi}} \int_0^\infty \int_0^{2\pi} p(\rho, x, t_0) e^{ik_y \rho \cos\theta} \rho \, d\theta \, d\rho,$$

(10)

where $\rho = \sqrt{y^2 + z^2}$ and $\theta = arctan(z/y)$. Since $p$ is independent of $\theta$, the integral in Eq. (10) is a Fourier-Bessel integral, which has the equivalent form:

$$f(x, k_y, t_0) = \frac{1}{\sqrt{2\pi}} \int_0^\infty p(x, \rho, t_0) J_0(k_y \rho) \rho \, d\rho,$$

(11)

with the zero-order Bessel function represented by $J_0$. This integral is a member of a Hankel transform pair with an inverse transform given by

$$p(x,\rho,t_0) = \frac{1}{\sqrt{2\pi}} \int_0^\infty \int_0^{2\pi} f(x,k_y,t_0) J_0(k_y\rho) k_y dk_y.$$

(12)

The integrand of Eq. (12) is obtained from experimental data using Eq. (7) and Eq. (8). Evaluation of this integral gives the desired result: the pressure field in space at a given time.

## III. Procedures

*A. Experiment*

A diagram of the experimental setup is presented in Fig. 2. A coherent light source is supplied by a 30-mW, 633-nm HeNe laser (L). The beam is guided through a positioner (P) and modulated at a frequency equal to the acoustic pulse repetition rate using an IntraAction (A0M -125H) acoustooptic modulator (M). After modulation, the beam is spatially filtered and expanded (BE), collimated (CM), and directed by flat mirrors (FM) through the water tank containing the acoustic field, which is propagating normal to the light. After interacting with the sound field, the diffracted light is focused using a converging mirror (CM). In the focal plane, the zero-order of light appears as a bright point on the z-axis. This zero-order is attenuated and shifted by a quarter wavelength using a custom thin film optical filter (SF).

Four circular optical filters of diameters d = 100μm, 200μm, 400μm, and 800μm, were constructed. Each of these filters consists of a 50nm layer of platinum (Pt) on top of a 204nm layer of silicon dioxide ($SiO_2$) deposited onto a transparent silicon wafer. The thickness was chosen using a refractive index value of 1.96 for Pt and a value of 1.54 for $SiO_2$ in order to produce a total phase shift of $\pi/2$ radians. Plasma etching was used along the circular edges of the platinum layer to assure a sharp interface. The range of diameters was selected to allow for the finite size of the zero-order of light.

After passing through this filter, the light is recorded in the image plane (CCD) of the collecting mirror. The optic field at this point is the inverse spatial transform of the field after passing through the filter, as indicated by Eq. (12). The light intensity data is fed into a Pentium PC from a 640X480 resolution CCD camera and frame grabber combination. These data are tomographically reconstructed using an efficient matrix-based algorithm that is a discrete approximation to Eq. (12), Eq. (8), and Eq. (7).

The acoustic signal is provided by a circular, 2.8 -cm diameter, broad-band piezo-ceramic transducer with a peak resonance at 500 kHz. The signal to the transducer is supplied by a Real Time Systems (3508) arbitrary waveform generator (AWG) through an ENI (240L) power amplifier to provide a peak volt age of approximately 200V across the transducer. The AWG signal input to the amplifier also serves as a trigger to the time-delayed optic pulse.

*B. Simulation*

Solutions to the linearized acoustic wave equation under axial symmetry may be expressed in terms of the time derivative of the convolution of the transducer displacement velocity $v(t)$ and an impulse response function, $h(x, y, z, t_0)$ ,[12] i.e.

$$p(\rho, z, t) = \frac{\partial}{\partial t}[v(t) \otimes h(\rho, z, t)].$$

(13)

Contributions from the back-surface of the piezoceramic are included. This is achieved by adding to the front -surface velocity a function that has the same shape, but is opposite in phase, has delayed time of flight and is reduced in amplitude.[10]

The phase-contrast apparatus provides a "snapshot" of the acoustic pressure, giving spatial dependency at a constant time, $p(x, y, z, t_0)$. In order to numerically calculate the pressure field in space, the time convolution in Eq. (13) was calculated for successive axial values $z$ at time $t_0$. The calculated acoustic pressure matrix was 45 X 200 with variable radial and axial step sizes in order to accommodate the varying waveform shapes.

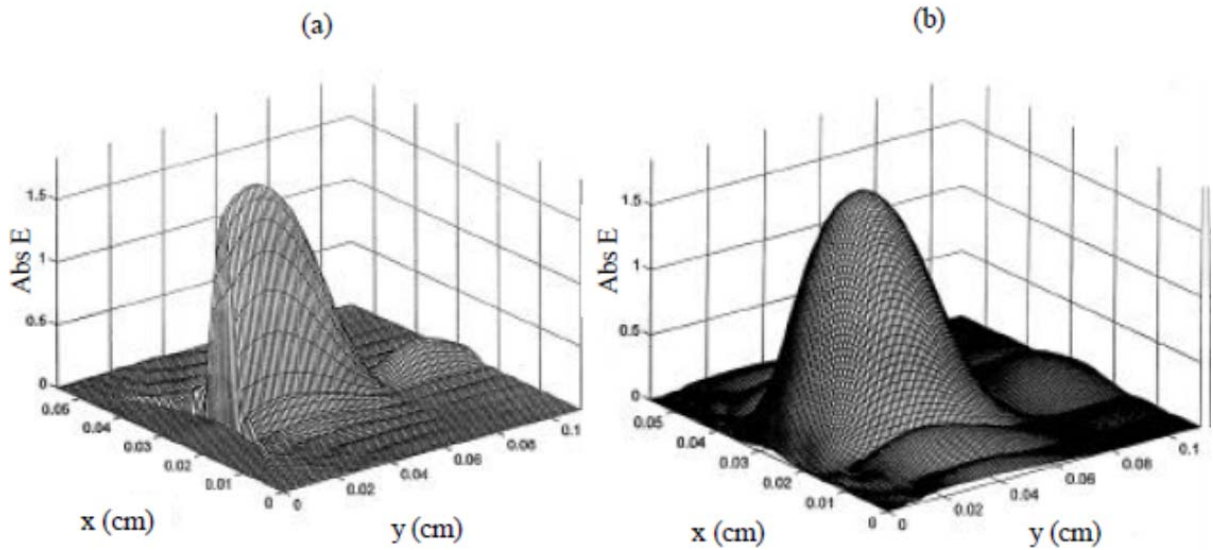

Figure 3. (a) Electric field intensity after passing through a 100μm phase-contrast filter. (b) Unfiltered Fourier plain after light is disturbed by an acoustic pulse. The zero-order of light has been removed.

## IV. Results and Discussion

The success of the phase-contrast method is dependent on its ability to shift the zero-order phase without disturbing the higher orders. Equation (6) represents the idealized situation where the unperturbed plane wave is focused to a point on the transform plane. Finite apertures and beam widths make this realization experiment ally impossible. Our procedure produces a beam of uniform intensity across a width of approximately 10cm. The minimum obtainable diameter for the undisturbed order in the focal plane is on the order of 10μm for the dimensions of the out lined optical apparatus. The finite size of the original spatial filter and alignment error make this dimension more realistically on the order of 100μm. A filter constructed with this diameter would shift this order and also any higher-order spatial frequencies that lie in this region. The effect of such a filter is illustrated in Fig. 3, showing the optic field intensity in the transform plane immediately after passing through the filter. The disturbing acoustic field is from a 1-MHz single-sine-cycle input into a 2.8 cm- diameter baffled planar disk measured at $t_0$ = 40 μs. The unfiltered plane is shown in Fig 3b with the zero-order (a delta function at the origin) removed.

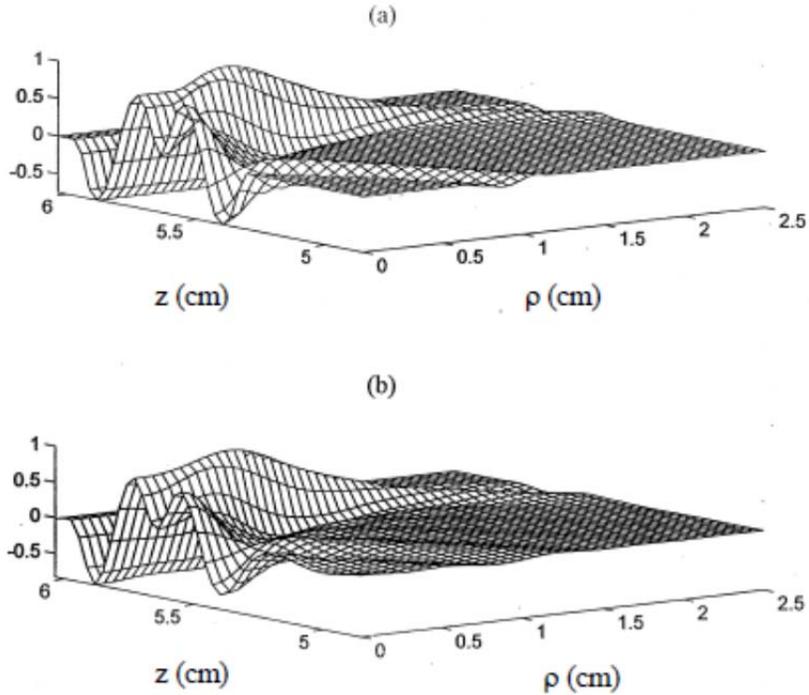

Figure 4. Simulated acoustic signal (a) at 40μs due to a 1MHz single-sine cycle input to a 1.4cm baffled planar disk and (b) Tomographic reconstruction of simulated signal after passing through an ideal phase-contrast filter.

Simulations indicate that if the spectral densities of the frequencies in this region are significant, the resulting image is distorted. The simulated acoustic signal at 40μs, using an impulse-response method, is presented in Fig. 4(a). The reconstructed signal after filtering only the zero-order is shown in Fig. 4(b). The signals after passing through 200μm, 400μm and 800μm filters are presented in Fig. 5. The reconstruction proves less accurate as the filter size increases. Similarly, the accuracy of the signal decreases with the frequency content of the signal. As in all acoustical imaging, the actual field is not known *a priori*, thus some calculation is necessary to estimate the valid filter diameter.

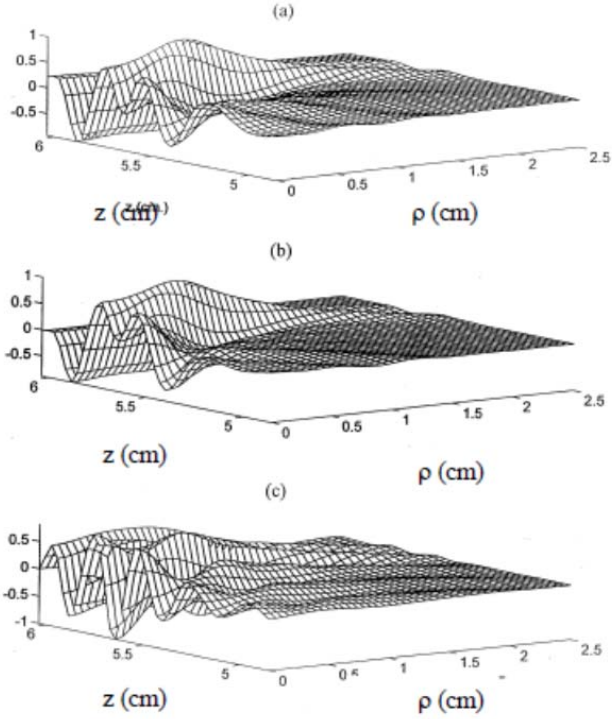

Figure 5. Reconstruction of simulated signal after passing through a (a) 200μm , (b) 400μm, and (c) 800μm -diameter filter.

Image quality dependence on the filter thickness was also investigated. Appreciable distortion of the image was found to occur only when the central order filter thickness caused a phase shift close to an integer multiple of π as expected from Eq. (5). Figure 6 shows a simulated field and reconstruction from a baffled planar source excited with a Gaussian input potential (full-width-at-half-maximum equal to 1μs) taken 18μs after the launching of the pulse. The field shown in Fig. 7 (a) is the tomographic reconstruction of the simulated field passed through the simulated 200μm–diameter filter illustrated in Fig. 6, but with an error in thickness causing a central order phase shift of π/6 beyond the intended π/2 shift. The figure indicates minor distortion in the field amplitude as a result of the thickness error. The distortion is more pronounced for an additional shift of π/3, displayed in Fig. 7 (b). An additional central order shift of π/2, presented in Fig. 7 (c), returns a field proportional to the absolute value of the pressure field upon reconstruction. This is similar to the reconstruction expected from a sensor using a central-order schlieren filter method.

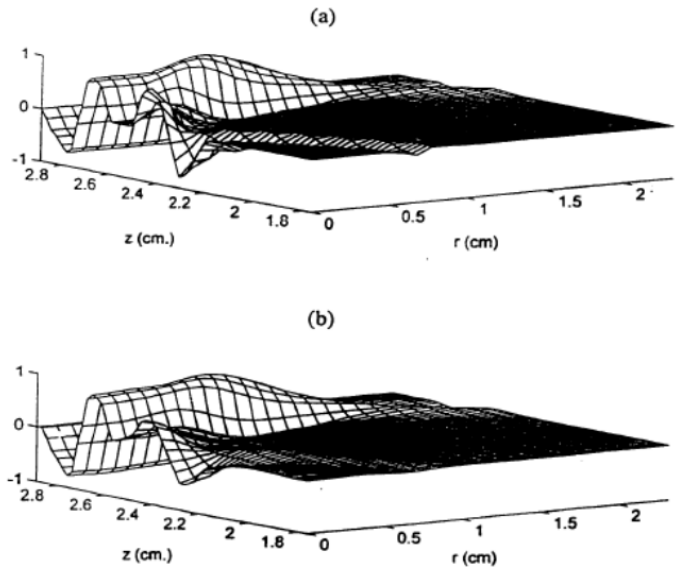

Figure 6. (a) Simulated acoustic signal 18μs after the launch o f a Gaussian input potential, (b) Tomographic reconstruction of simulated signal after passing through an ideal phase-contrast filter.

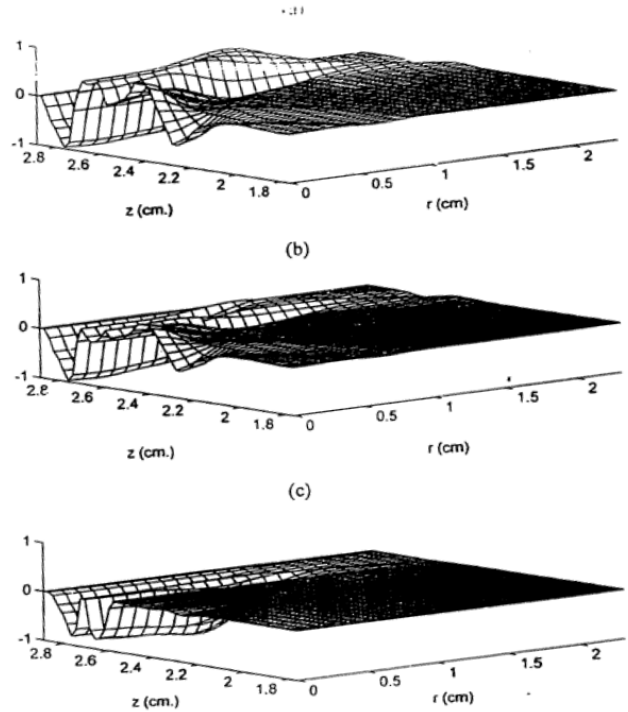

Figure 7. Reconstruction of simulated Gaussian field using a 200μm-diameter filter with an added phase error of (a) π/6, (b) π/3, and (c) π/2.

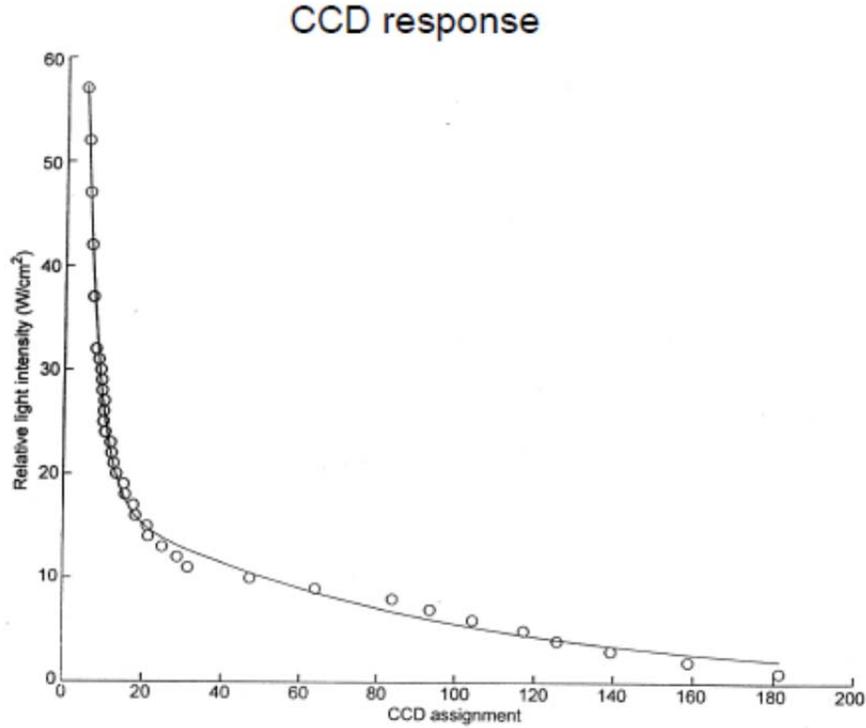

Figure 8. CCD response as a function of input light intensity.

To provide accurate reconstruction of the field, measurement s of the CCD were collected by a digit al frame grabber card (VideoSpigot) and assigned numeric values between 1 and 255 corresponding to relative light intensity. Lower numbers represent higher intensity. Controlled intensity measurements were made to calibrate the detector. The results, shown in Fig. 8, indicate that the detector is nonlinear at higher intensities, as it approaches its saturation point. To scale measurements, the light intensity data were fit to a function of the form

$$I = C_1 e^{-\lambda_1 I_D} + C_2 e^{-\lambda_2 I_D}$$

(14)

Where $I_D$ are the measured data, $I$ are the scaled data, $C_1$ and $C_2$ are linear scaling parameters and $\lambda_1$ and $\lambda_2$ are nonlinear scaling parameters. Using a Matlab (1994, The Mathworks, Inc.) fitting function, the parameters were found to have the following values: $C_1$=148.83, $C_2$=18.36, $\lambda_1$=0.2710, $\lambda_2$= 0.0119. All measured field data have been scaled using these parameters.

To demonstrate the phase-contrast method, a Gaussian driving signal with peak volt age of 200V and full-width-at-half-maximum of 1μs is generated by the AWG, input to the transducer, and imaged. Based on simulations, the 200μs Fourier filter was chosen for the imaging. An example of the intensity images recorded by the system is presented in Fig. 9.

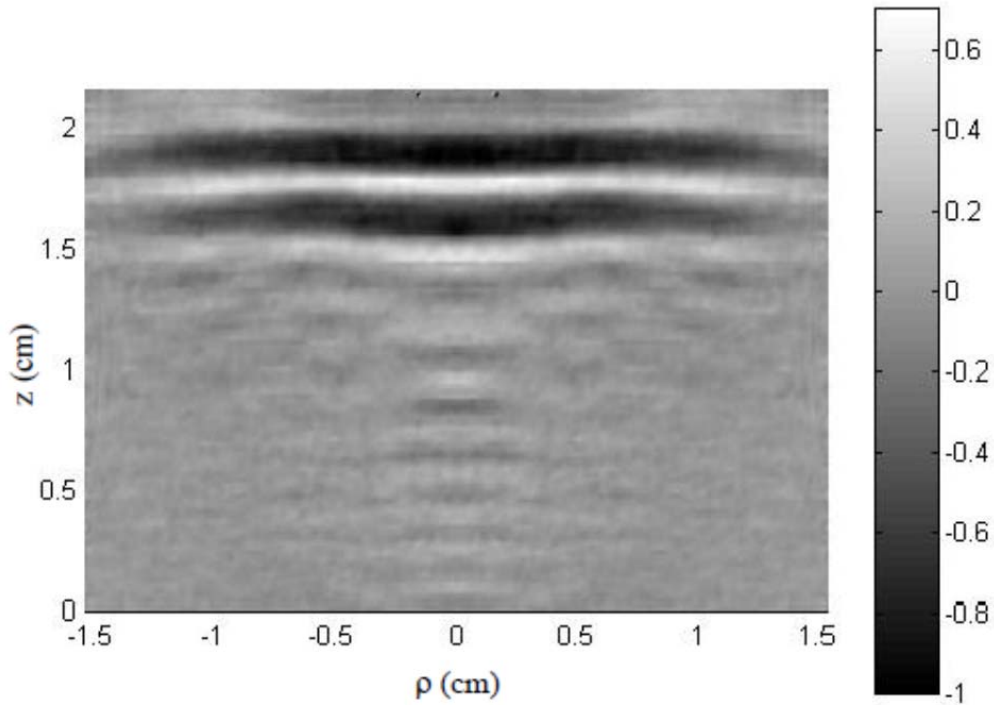

Figure 9. Example of a raw image of an acoustic pulse acquired with the phase-contrast approach, demonstrating the gray background and effects due to both positive and negative pressure. The pulse is traveling in the positive z-direction.

Each intensity value in the 100X 200 matrix is assigned a numerical value in the range -127 to 127, zero representing the average background light intensity. The data are next scaled using Eq. (14). The matrix is then input into the reconstruction algorithm, yielding the pressure field in space at a fixed time $t$. The reconstructed pressure fields at 4μs, 16μs, and 22μs are presented in Fig. 10. These measured fields are compared with a numerically calculated field in Fig. 11.

The physical signals in Fig. 10 contain more oscillations along the z-direction than the ideal case. Additionally, the on-axis peak occurring at 1.8cm in the simulated 16μs signal appears at 2cm in the real data, closer to the main portion of the signal. In general, the real data decays faster in the radial direction. Certain general features appear both in the simulation and the experiment, including the arcing of the signal in the radial direction along the peak at 2.2cm.

## V. Summary

The field due to an underwater, ultrasonic source is imaged using an optical phase-contrast method. Relevant theory is out lined along with a description of the apparatus. The study uses a straight -ray approach in the Raman-Nat h region and concentrates on the effect s of signal frequency content on the filter. Special attention is noted regarding the dimensions of the phase-contrast Fourier filter dimensions. It is demonstrated that the appropriate filter diameter is dictated by its ability to shift the zero-order (undiffracted) of light without significantly affecting higher-orders. Acoustic signals with high temporal frequency content will generally display high spatial frequencies and thus are accurate even for spatial filters with relatively large diameter. For a 1 MHz sine-pulse, a 200μm filter is found to be sufficiently large to phase shift the zero-order without distorting the higher orders. However, larger filters are shown to distort the reconstructed acoustic signal.

The basic purpose of the phase-contrast filter is to shift the zero-order of light in phase by $\pi/2$ while attenuating it by at least 90%. Simulations are performed to examine the sensitivity of the phase-contrast method to the phase shift by simulating error in filter thickness. It was concluded that a sensor would be relatively insensitive to error in filter thickness causing a light shift of less than $\pi/6$. For a glass thin- film layer this would allow a thickness error of over 90nm.

The phase-contrast apparatus is demonstrated by reconstructing the field due to a low-intensity Gaussian potential input to an underwater ultrasonic source. Comparison of the measured field with numerical simulation suggests reasonable results. The phase-contrast method provides the same advantages as traditional schlieren imaging including rapid, noninvasive data acquisition and imaging in regions where a receiver may cause interference. Additionally, the method measures the instantaneous acoustic pressure field, where schlieren imaging returns a signal proportional to the modulus of the field thus providing a tool for imaging the acoustic transient waveforms in space at a given time.

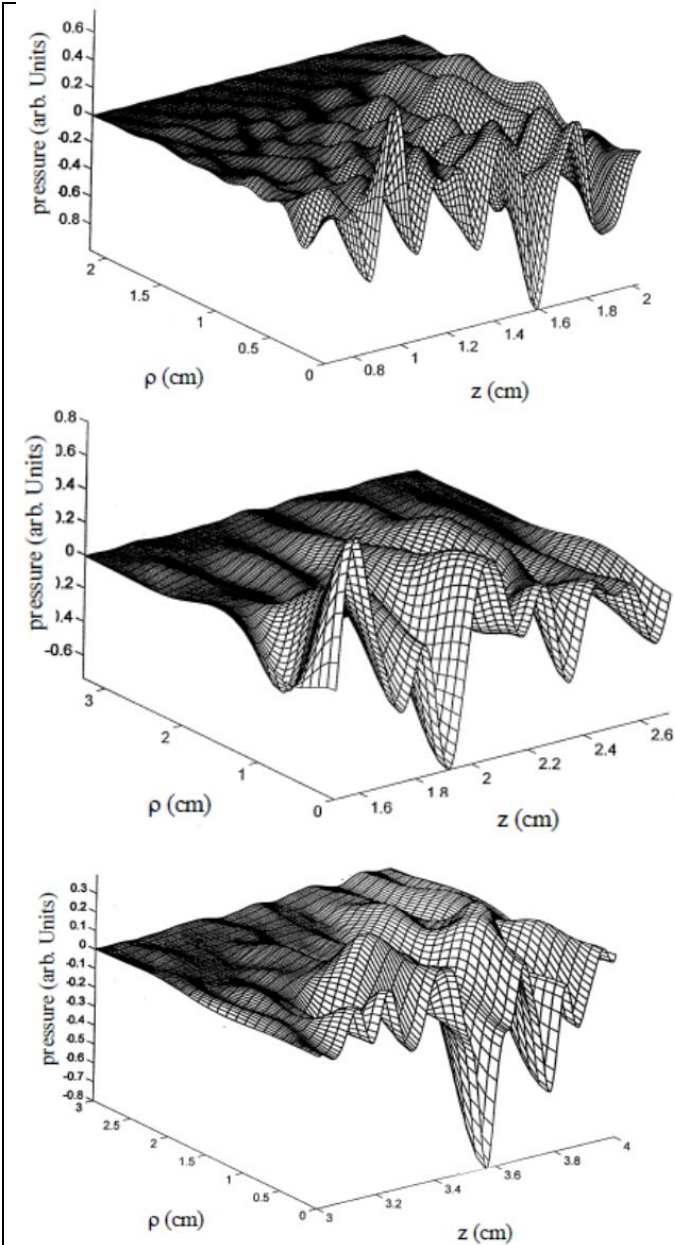

Figure 10. Tomographic reconstructions of pressure fields at 4µs, 16µs, and 22µs.

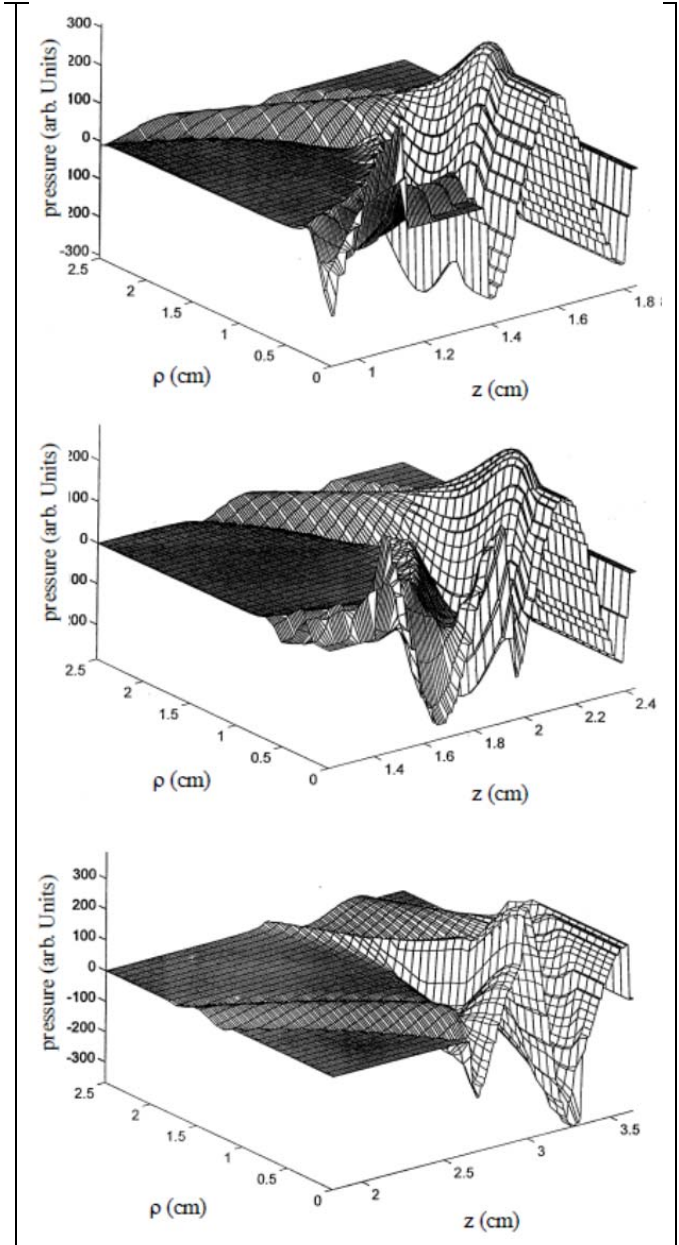

Figure 11. Simulations of pressure fields at 4µs, 16µs, and 22µs.

18